\documentclass[onecolumn]{emulateapj}  
\usepackage[pagewise]{lineno}

\lefthead{Wada}
\righthead{CO lines in Circinus}

\newcommand{\bea}{\begin{eqnarray} }
\newcommand{\eea}{\end{eqnarray}}

\begin{document}


\title{Circumnuclear Multi-phase Gas in Circinus Galaxy I:  Non-LTE calculations of CO lines}
 
\author{%
Keiichi Wada
}
\affiliation{Kagoshima University, Graduate School of Science and Engineering, Kagoshima 890-0065, Japan}
\affiliation{Ehime University, Research Center for Space and Cosmic Evolution, Matsuyama 790-8577, Japan}
\affiliation{Hokkaido University, Faculty of Science, Sapporo 060-0810, Japan}
\email{wada@astrophysics.jp}

\author{
Ryosuke Fukushige
}%
\affiliation{Kagoshima University, Graduate School of Science and Engineering, Kagoshima 890-0065, Japan}

\author{Takuma Izumi}%
\affiliation{National Astronomical Observatory of Japan, Mitaka 181-8588, Japan}
\affiliation{NAOJ Fellow}

\author{Kohji Tomisaka}
\affiliation{National Astronomical Observatory of Japan, Mitaka 181-8588, Japan}

%


\begin{abstract}
In this study, we investigate the line emissions from cold molecular gas based on our previous ``radiation-driven 
fountain model'' \citep{wada2016}, which reliably explains the spectral energy distribution of the nearest 
type 2 Seyfert galaxy, the Circinus galaxy. Using a snapshot of the best-fit radiation-hydrodynamic model for 
the central $r \le 16$ pc, in which non-equilibrium X-ray-dominated region chemistry is solved,  
we conduct post-processed, non-local thermodynamic equilibrium radiation transfer simulations
for the  CO lines. We obtain a spectral line energy distribution
with a peak around $J\simeq 6$, and its distribution suggests that the lines are not thermalized.
However, for a given line-of-sight, the optical depth distribution is highly non-uniform between $\tau_\nu \ll 1 $ and $\tau_\nu \gg 1$.
The CO-to-H$_2$ conversion factor ($X_{\rm CO}$), which can be directly obtained from the results, is not
a constant and depends strongly on 
the integrated intensity, and
it differs from the fiducial value for local objects.
$X_{\rm CO}$ exhibits a large dispersion of more than one order of magnitude, reflecting the non-uniform internal structure of a ``torus.'' 
 We also found that  the physical conditions differ between grid cells on a scale of a few parsecs along the observed lines of sight; therefore, a specific observed line ratio does not necessarily represent a single physical state of the ISM.  
\end{abstract}


\section{INTRODUCTION}
%

In the standard picture of active galactic nuclei (AGNs), the nucleus is hypothetically 
surrounded by an optically and geometrically thick ``dusty torus.'' 
Except for a few nearby AGNs, such as NGC 1068, NGC 1097 and the Circinus galaxy {\citep[e.g.,][]{jaffe04, tristram2014, garciaburillo2016, imanishi2016, gallimore2016, izumi2017}}, dust and molecular emissions have not yet been spatially resolved; 
therefore, the real geometrical and internal structures of the tori are still unclear. 
The origin and physical mechanism of the obscuring material around the nucleus 
have been widely discussed by many authors \citep[e.g.,][]{kro88, pier93, ohs01, lawrence2010, hopkins2012}.
Recent time-dependent radiation-(magneto)hydrodynamic simulations
commonly suggested that a static, donut-like torus is not reproduced, {because on
sub-parsec to tens of parsec-scales, the ISM is very dynamic} \citep{wada2012, wada2016, namekata2016, dorodnitsyn2012, dorod2016, dorod2017, chankrolik2016,chankrolik2017}, although there are non-negligible differences in the results among the simulations.

\citet{wada2012} proposed that the obscuring structures around 
AGNs, in which outflowing and inflowing gases are driven by 
radiation from the accretion disk, form 
a geometrically thick disk on the scale of a few parsecs to tens of parsecs.
The quasi-steady circulation of gas, i.e., the ``radiation-driven fountain,'' may
 obscure the central source, therefore, 
  the differences in the spectral energy distributions (SEDs) 
 of typical type 1 and 2 Seyfert galaxies
 are reliably explained \citep{schartmann2014}.
\citet{wada2015} showed that the observed properties of obscured AGNs
change as a function of their luminosity because of fountain flows, and the results were compared
with recent X-ray and infrared observations \citep[see also][]{almeida2017}.

\citet{wada2016} (hereafter W16) applied this radiation-driven fountain model to 
the Circinus galaxy, which is the nearest ($D=$ 4 Mpc) type-2 Seyfert galaxy.
We studied, for the first time, the non-equilibrium chemistry for the X-ray-dominated region (XDR)
with supernova feedback in the central $r \le 16$ pc\footnote{{The photon dominated region (PDR) chemistry and shock chemistry are not
considered in the present model.}}.
A double hollow cone structure occupied by an inhomogeneous, diffuse ionized gas is formed, 
 and it is surrounded
by geometrically thick ($h/r \gtrsim1$) atomic/molecular gas.
Dense molecular gases {(e.g. H$_2$ and CO with $n_{{\rm H}_2} \gtrsim 10^3$ cm$^{-3}$)} are mostly concentrated around
the equatorial plane, and {atomic gas (e.g. H$^0$ and C$^0$)} extends with a larger scale height.
The energy feedback from supernovae enhances its
scale height.
In W16, by applying post-processed three-dimensional (3-D) radiation transfer calculations, we found 
``polar'' emission in the mid-infrared band (12 $\mu {\rm m}$), which is associated with
bipolar outflows, as suggested in recent interferometric observations of nearby AGNs \citep[][see also \citet{stalevski2017} for a theoretical model]{hoenig2013,tristram2014}.
We also confirmed that the viewing angle $\theta_v$ for the nucleus should be larger than 75$^\circ$ (i.e., close to edge-on) in order to
explain the observed SED and 10 $\mu m$ absorption feature of the Circinus galaxy \citep{prieto2010}.

In this study, we focus on the structures of the cold, molecular gas located in the outskirts of
the infrared bright region in the best-fit model for the Circinus galaxy.
In order to compare with future high-resolution observations of the molecular gas, 
we performed 3-D, non-local thermodynamic equilibrium (non-LTE) radiative transfer 
calculations for the $^{12}$CO lines. Spatial distributions of the multi-$J$ line intensity, line ratios, spectral line energy distribution (SLED), and CO-to-H$_2$ conversion factor are discussed
in the central ten parsecs with a sub-parsec resolution.
The results can be a theoretical reference to understand observations of the Atacama Large Millimeter/submillimeter Array (ALMA), whose spatial resolution can now reach a few parsecs in nearby AGNs.
 Detailed comparisons between the models and our Cycle-4 observations by ALMA 
 will be discussed in a subsequent paper (Izumi et al., in preparation).
The results of other molecules (e.g., HCN) and atomic gases (e.g., neutral carbon and hydrogen) will be discussed elsewhere.

\section{NUMERICAL METHODS AND MODELS}
%

\subsection{Input model:  Radiation-driven fountain}
We use a snapshot of our radiation-driven fountain model, which reliably explains the
features of the SED of the Circinus galaxy (see details in W16).  
The results suggest that the viewing angle for the gas disk should be 75$^\circ$ or higher.
We use our 3-D Eulerian hydrodynamic code \citep{wada2012, wada2015}, 
 with a uniform grid that accounts for radiative feedback processes from the central source using a ray-tracing method.
We include the non-equilibrium XDR chemistry \citep{maloney96,
meijerink05} for 256$^3$ zones (resolution of 0.125 pc). 
Self-gravity of the gas is ignored because it is not essential for the gas dynamics in the radiation-driven fountain (see also \citet{namekata2016}). Supernova feedback is also implemented.
Cooling functions for 20 K $\le T_{gas} \le 10^{8}$ K \citep{meijerink05, wada09} and 
solar metallicity are also assumed.
At every time step, the gas density, $T_{gas}$ and $T_{dust}$\footnote{If the temperature of the dust irradiated by a central source $T_{dust}$ exceeds $1500\,{\rm K}$, we assume the dust is sublimated, 
and no dust is assumed for $T_{gas} > 10^5 $ K, because of dust sputtering. }, and ionization parameters in the 256$^3$ grid cells
are passed to the chemistry module. 

The central point mass is assumed to be $2\times 10^6 M_\odot$,
which is comparable to the value estimated from the maser observations \citep{greenhill2003}, 
$1.7\pm 0.3 \times 10^6 M_\odot$ in Circinus. The total gas mass is $2\times 10^6 M_\odot$.
%
The Eddington ratio of 0.2 and the bolometric luminosity of $L_{bol}=5 \times 10^{43}$ erg s$^{-1}$ 
are fixed during the calculation. 

{Here we assume  three independent  radiation fields; 1) non-spherical UV radiation emitted from
a thin accretion disk,  2) spherically symmetric X-ray radiation emitted from the corona of the accretion disk  \citep{netzer1987, xu2015}, 
and 3) uniform far UV due to star forming regions exisited in the circumnuclear disk. All these three components are assumed to be time-independent.
The first component plays an important role for the radiation pressure to the dust, 
and the second one mainly contributes for heating the gas depending on the column density, and also for
the non-equilibrium XDR chemistry\footnote{We use a selection of reactions from the chemical network described by \citet{meijerink05, adamkovics2011}
for 25 species:  H, H$_2$, H$^+$, H$_2^+$, H$_3^+$, H$^-$, e$^-$, O, O$_2$, O$^+$, O$_2^+$, O$_2$H$^+$, OH,
OH$^+$, H$_2$O, H$_2$O$^+$, H$_3$O$^+$, C, C$^+$, CO, Na, Na$^+$, He, He$^+$, and HCO$^+$.}.
The SED of the AGN and the dust absorption cross section are taken from \citet{laor1993}.
The UV flux is assumed to be  angle dependent, i.e. $F_{UV}(\theta) \propto \cos \theta (1+2\cos \theta)$ \citep{netzer1987},
where $\theta$ denotes the angle from the rotational axis ($z$-axis), which
is calculated for all grid cells using $256^{3}$ rays. 
The UV and X-ray fluxes are calculated from the bolometric luminosity \citep{marconi2004}. 
The total X-ray luminosity (2--10 keV) is $L_X = 2.8\times 10^{42} $ erg s$^{-1}$. 
The temperature of the interstellar dust $T_{dust}$ at a given position irradiated by
the central UV radiation is calculated by assuming thermal equilibrium \citep[e.g.,][]{nenkova2008}, 
therefore, it is not necessarily equivalent to $T_{gas}$.
}

{For simplicity, the third component of the radiation field, i.e. FUV, is assumed to be uniform.
In stead of solving the radiation transfer for FUV in the inhomogeneous media with multi-radiation source, 
which is beyond our numerical treatment, we change its strength from $G_0=100$ to $G_0 = 10^4 $,
where $G_0$ is the incident FUV field normalized to the local interstellar value ($1.6\times 10^{-3}$ erg cm$^{-2}$ s$^{-1}$), 
and see if it affects the results (see \S 4.2.)}.

The 3-D hydrodynamic grid data (i.e., density, temperature, abundances, and three components of velocity) of the 256$^3$ grid cells
are averaged to produce 128$^3$ grid cells (i.e., the spatial resolution is 0.25 pc) in order to 
reduce the computational cost. This is passed to the 3-D non-LTE line transfer code, as described in the next section, to
derive the line intensities.


\subsection{Non-LTE line transfer}
The numerical code of the 3-D line transfer calculations is the same as that used by \citet{wada05} and \citet{yamada07}, which is based on the Monte--Carlo and long-characteristic transfer code \citep{hoge00}.  The statistical equilibrium rate equations and the transfer equations are iteratively solved
using photon packages (sampling rays) propagating into each grid cell. We solve the rate equations for the energy levels of $^{12}$CO from $J = 0$ to 15. Once the radiation field and optical depth are determined for all grid cells, we ``observe'' it from 
arbitrary directions, and 3-D data cubes (i.e., positions and line-of-sight velocity) are obtained for selected transitions.


There are 1000 sampling rays for each grid cell. The level populations are converged 
with an error of $10^{-6}$ for $J=1$ and $10^{-3}$ for $J=4$ after ten iterations. 
The micro-turbulence, i.e., a hypothetical turbulent motion inside one grid cell, is a free parameter,
which determines the shape of the line profile function and varies from $v_{turb} =$1 to 20 km s$^{-1}$.
As a fiducial value, we assume $v_{turb} = 10$ km s$^{-1}$, and the effect of changing $v_{turb}$ is discussed in \S 4.
In contrast to previous papers \citep{wada05, yamada07}, 
we use here the non-uniform abundance distribution for $^{12}$CO, 
which is obtained in the original radiation-hydrodynamic simulations with XDR chemistry (W16). 
{Note that $v_{turb}$ is different from the typical velocity dispersion on several pc scales.
In fact, the average velocity dispersion of the gas in the input model on a few pc is about 30 km s$^{-1}$, which 
is consistent to the scale height of the cold ($< 100$ K), molecular gas seen in Fig. 1. 
In the turbulent gas disk, the velocity dispersion is larger on larger scale as expected from 
the power-law energy spectrum \citep[e.g.][]{wada2002}. Therefore we expect that the velocity dispersion inside
a grid cell should be smaller than the large scale velocity dispersion.
The ``large scale" bulk motion between grid cells in the disk is self-consistently considered in the non-LTE 
line transfer calculations.
}

\begin{figure}[h]
\centering

\includegraphics[width = 13cm]{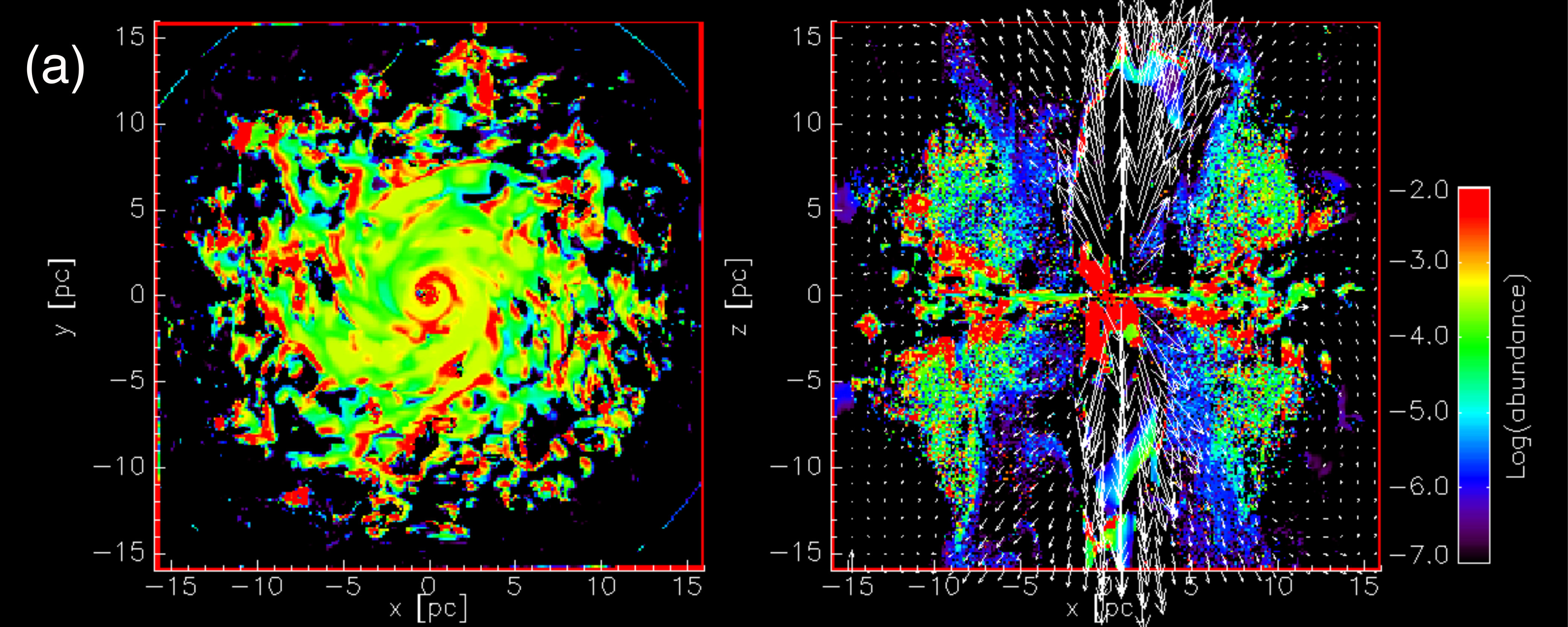}
\includegraphics[width = 8cm]{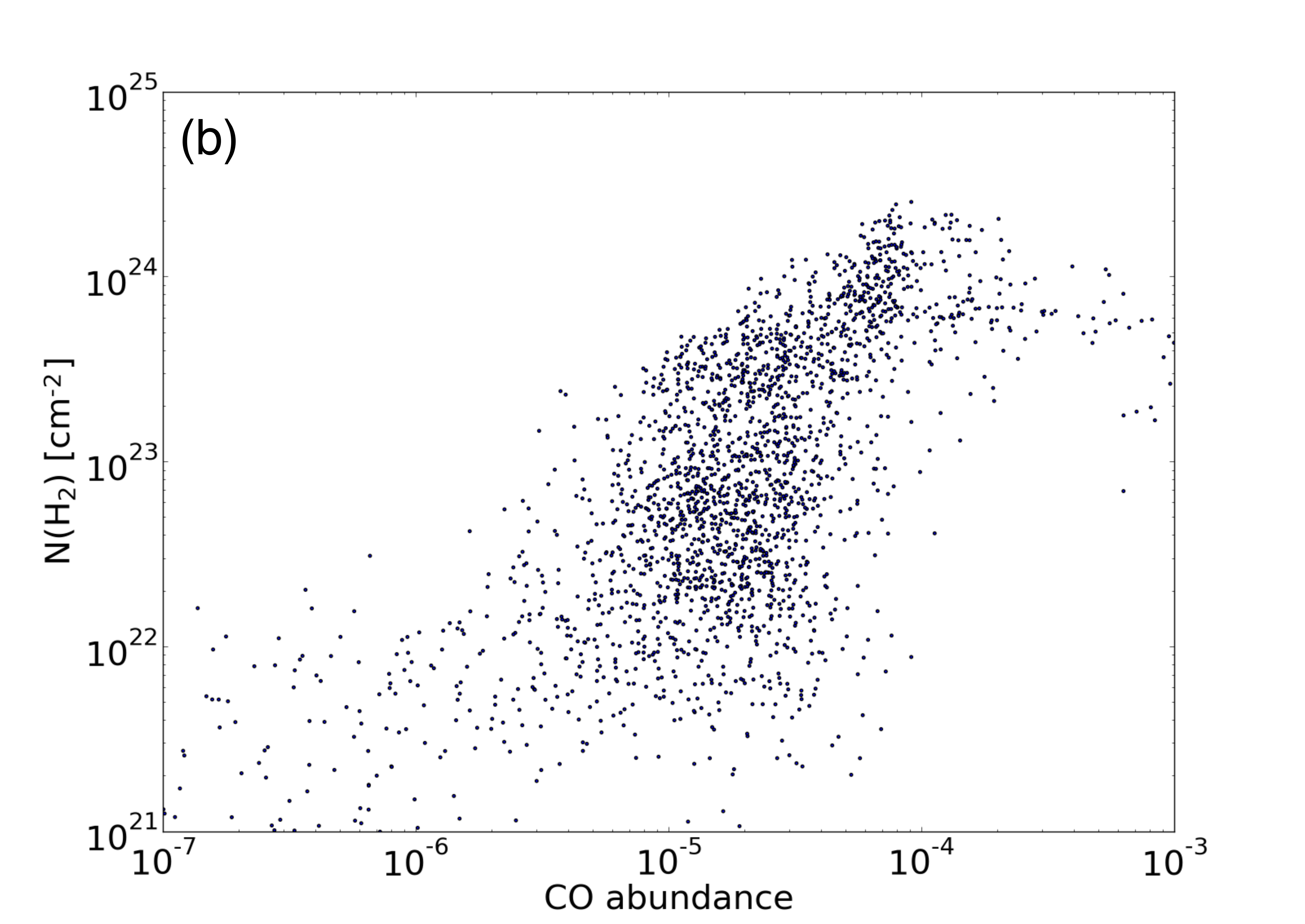}

\caption{(a) Abundance distribution of CO on $z=0$ and $y=0$ pc planes in an input model with FUV of $ G_0=10^3$. (b) Abundance as a function of H$_2$ column density. }
\label{wada_fig: 1}
\end{figure}

Figure \ref{wada_fig: 1}a presents spatial distributions of the $^{12}$CO abundance with respect to molecular hydrogen in the input model assuming $G_0= 10^3$.  
As shown in Fig. \ref{wada_fig: 1}b, the $^{12}$CO abundance is $x_{\rm CO} = 10^{-5}-10^{-4}$ for most grid cells,
but there is large scatter for a given column density of the inhomogeneous disk. 
Hereafter, ``CO'' represents $^{12}$CO.

%
%
%
\section{RESULTS}
%

\subsection{Intensity distribution}

\begin{figure}[h]
\centering
\includegraphics[width = 14cm]{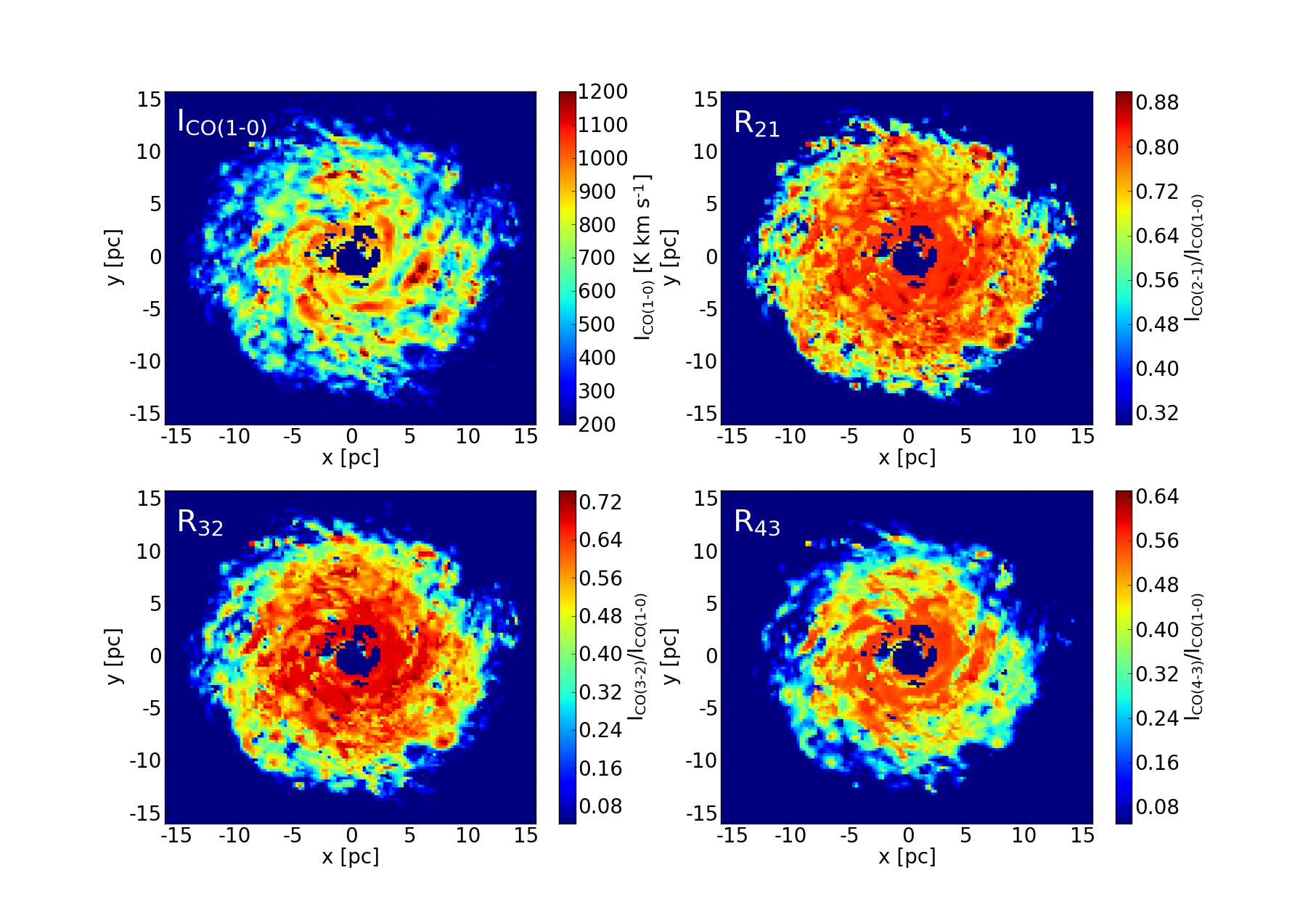} 

\caption{Integrated intensity map of CO(1-0) and 
line ratio distributions ($I_{\rm CO2-1}/I_{\rm CO1-0}$, $I_{\rm CO3-2}/I_{\rm CO1-0}$, and $I_{\rm CO4-3}/I_{\rm CO1-0}$) for $\theta_v = 0^\circ$ (face-on). 
 }
\label{wada_fig: 2}
\end{figure}

\begin{figure}[h]
\centering
\includegraphics[width = 14cm]{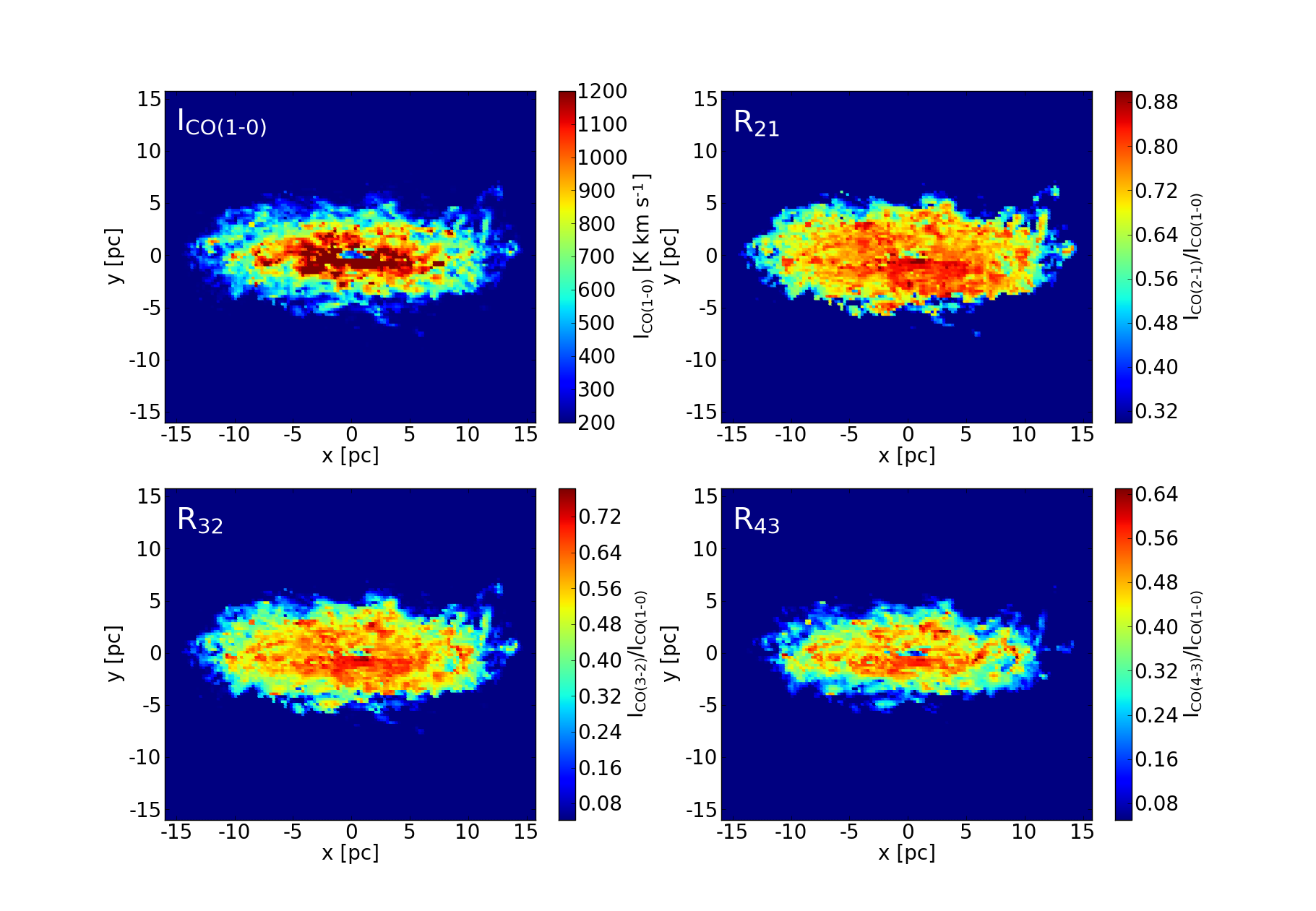}  
\caption{Same as Fig. \ref{wada_fig: 2}, but for $\theta_v = 75^\circ$, which is 
 is suggested from the fit with the infrared SED of the Circinus galaxy (W16). }
\label{wada_fig: 2b}
\end{figure}

Figures \ref{wada_fig: 2} and  \ref{wada_fig: 2b} show the integrated intensity maps of CO (1-0) and the line ratio distributions ($J=$ 2-1/1-0, 3-2/1-0, and 4-3/1-0) in K km s$^{-1}$
for the viewing angles $\theta_v = 0^\circ$ (i.e., face-on) and 75$^\circ$, repspectively. The intensity distributions are morphologically similar between the lines. The face-on maps exhibit 
multi-arm spiral-like features with cloud-like high-density regions. Inhomogeneous structures are also prominent from $\theta_v = 75^\circ$. In the following results, $\theta_v =75^\circ$, which is
suggested by comparing with the SED of the Circinus galaxy (W16), is assumed, if not stated. 
{We also notice that the most CO emissions originate from the disk region, and there are no dense ``molecular winds'' outflowing
along the rotational axis.  This would be interesting to understand 
the origin of the kpc-scale molecular outflows observed in nearby AGNs, such as NGC 1068\citep{garciaburillo2014} (see also discussion in \S 4.3).}

%

%

%
%

\begin{figure}[h]
\centering
\includegraphics[width = 9cm]{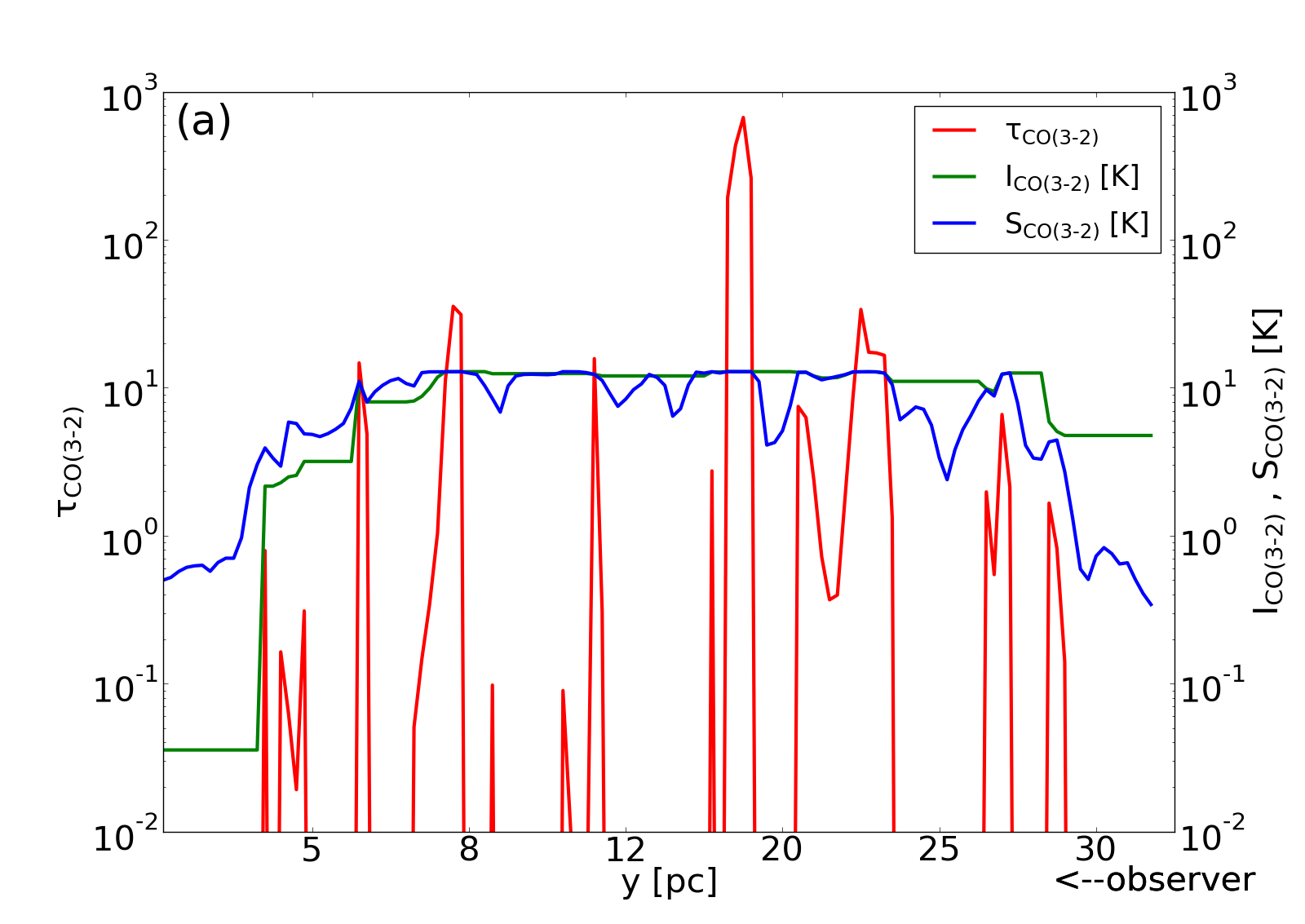}  
\includegraphics[width = 9cm]{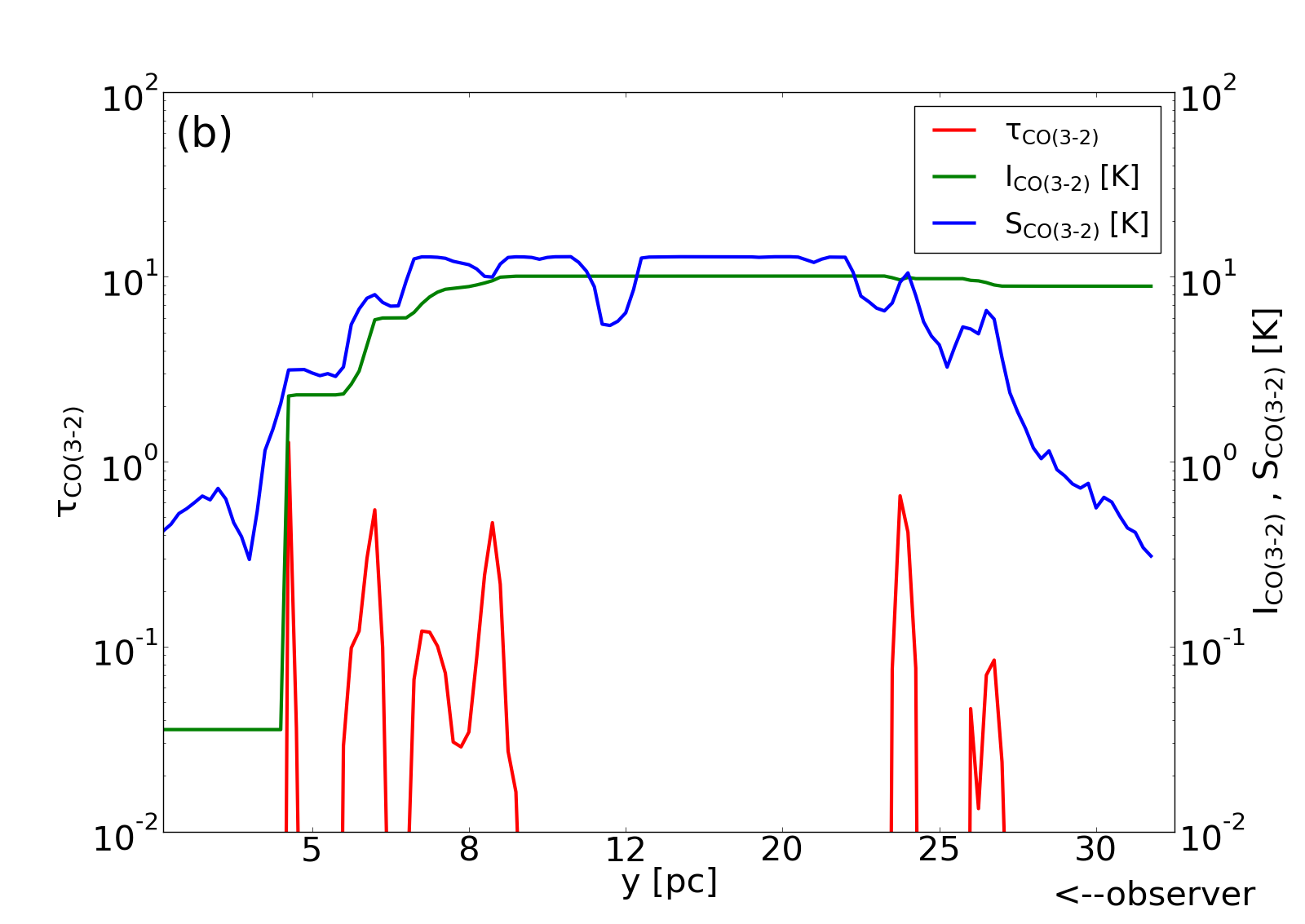}  
\caption{Optical depth ($\tau_{\rm CO(3-2)}$), intensity ($I_{\rm CO(3-2)}$), and source function ($S_{\rm CO(3-2)}$) in each grid cell along two line-of-sight directions selected from relatively thick and thin regions. The ray is transferred from the left-most grid cell (i.e., the background radiation) to the right-most grid cell, where the observer is located. For both plots, a central velocity channel around the systematic velocity, i.e., $ 2.5 \; {\rm km} \; {\rm s}^{-1} < |v| $, is used. }
\label{wada_fig: fig3}
\end{figure}

In Fig. \ref{wada_fig: fig3}, the changes in the intensity $I_\nu$, source function $S_\nu$, and optical depth $\tau_\nu$
of CO (3-2) along two different rays toward the observer in a velocity channel ($\pm 2.5$ km s$^{-1}$ around the systematic velocity) are shown.
Along the rays, the optical depth $\tau_\nu$ spatially changes by many orders of magnitude. There are 
several optically thick regions, which 
are non-uniformly distributed, and they form several discrete ``clouds'' along this sample line-of-sight (Fig.  \ref{wada_fig: fig3}a).
The intensity increases from the background (i.e., CMB) at $ y \sim 3-10$ pc,
where $0.1 \lesssim \tau_\nu $, and is saturated at the clouds where $\tau_\nu > 1$ or stays constant
if $\tau_\nu \ll 1$. At the observer side ($y \gtrsim 25$ pc), the intensity exceeds the source function and is therefore absorbed at the nearside grids depending on the local optical depth.
A relatively optically thin case is also shown in Fig. \ref{wada_fig: fig3}b.
The intensity increases following $\sim \tau_\nu S_\nu$ at the far side (i.e., $y \lesssim 10$ pc), and it is slightly absorbed at the near side.
This represents the characteristic feature of the line transfer effect through a highly non-uniform medium.
It is then inferred that the observed integrated intensity in terms of the line-of-sight velocity does not necessarily represent a {\it single} state of the internal structure of the
circumnuclear disk along a given line-of-sight (see also \S 3.4).

%


%
\subsection{X-factor}
%
As it is difficult to directly observe emission from cold molecular hydrogen,
the line intensity of CO (often $J=1 \rightarrow 0$) is used to
estimate the column density of H$_2$ in giant molecular clouds in our Galaxy, as well as the total molecular mass in
external galaxies. The CO-to-H$_2$ conversion factor is often represented 
as $X_{\rm CO}$ for the number density or $\alpha_{\rm CO}$ for the mass density.
Based on various independent methods, it is estimated for the Galactic disk that $X_{\rm CO(1-0)} \simeq 2\times 10^{20}$ cm$^{-2} ({\rm K}\; {\rm km}\; {\rm s}^{-1})^{-1}$
or $\alpha_{\rm CO(1-0)} = 4.3 M_\odot \; {\rm pc}^{-2}  ({\rm K}\; {\rm km}\; {\rm s}^{-1})^{-1}$, using the CO(1-0) intensity,
and the error could be within a factor of 2--3 for molecular clouds in our Galaxy \citep[][]{bolatto2013}.
Although this fiducial value is often also used for observations of external galaxies,  
it should not be the same in different environments, such as the ISM around AGNs or in 
starburst regions.
 Here, we present an independent estimate of the conversion factor using the line transfer calculations shown in 
 \S 3.1. One should note that the values could be the case for the central region of the Circinus galaxy 
 and could also be applied to low-luminosity AGNs associated with
 nuclear starbursts, but it is not necessarily the case for all types of AGNs, such as luminous quasars.

\begin{figure}[h]
\centering
\includegraphics[width = 16cm]{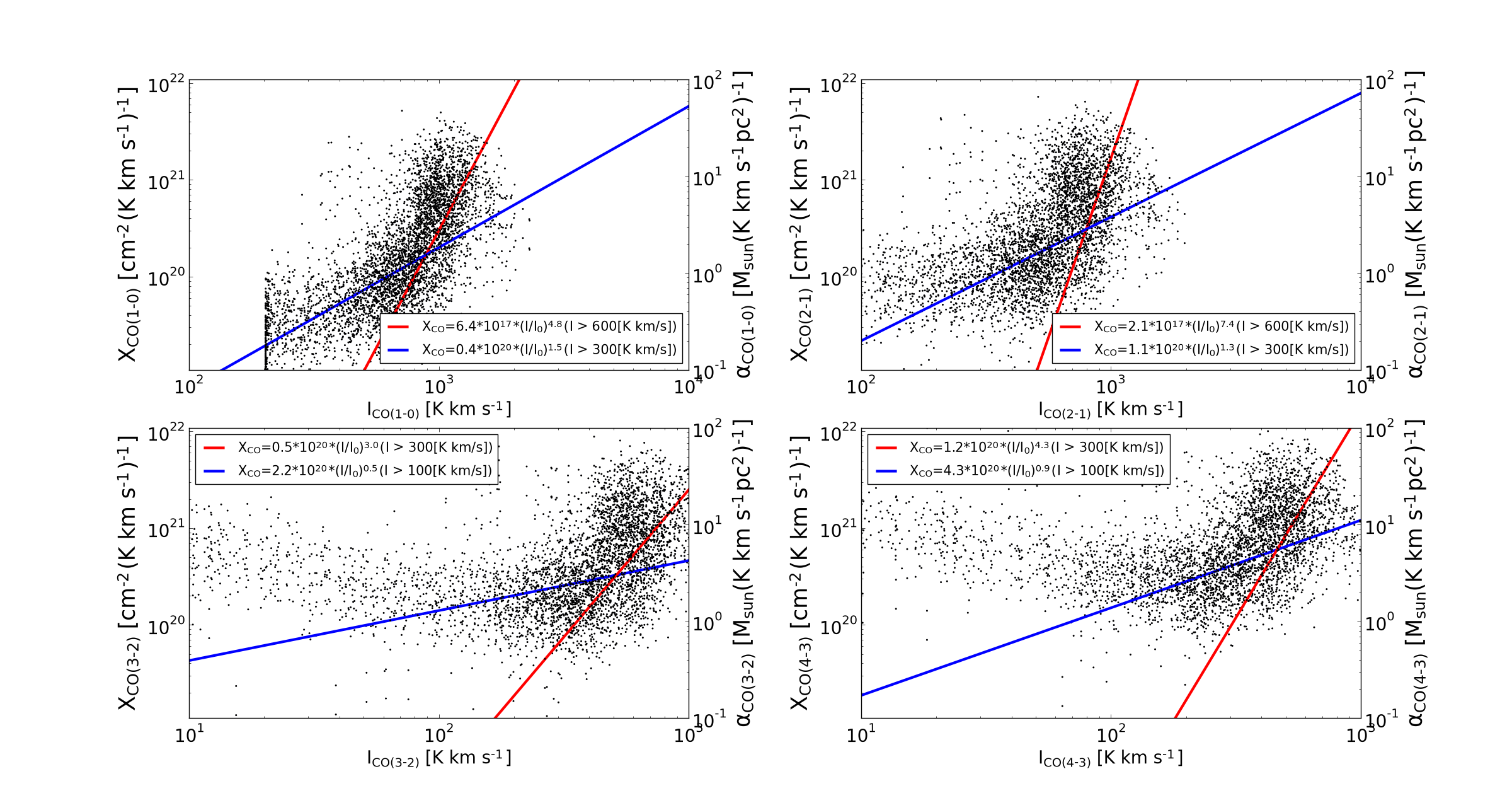} \\  
\caption{CO-to-H$_2$ conversion factor for four lines. The inclination angle is 75$^\circ$. }
\label{wada_fig: 4}
\end{figure}

Figure \ref{wada_fig: 4} shows X$_{\rm CO}$ (and $\alpha_{\rm CO}$ for the right vertical axis) 
as a function of the line intensities
of CO(1-0), CO(2-1), CO(3-2), and CO(4-3) based on the non-LTE line intensity calculations and H$_2$ density in the input data.
It is apparent that the CO-to-H$_2$ conversion factor is not a constant and
strongly depends on the line intensity, especially for the low-$J$ lines.
If we fit the conversion factor for $J=1-0$ with a single power law, we can obtain
$X_{\rm CO(1-0)}  \simeq 2.0\times 10^{20}$ cm$^{-2} (I_{\rm CO(1-0)}/I_0)^{4.8}$  $({\rm K}\; {\rm km}\; {\rm s}^{-1})^{-1}$, or
$\alpha_{\rm CO} \simeq 4.3 M_\odot \;  (I_{\rm CO(1-0)}/I_0\; )^{4.8}$  $({\rm K}\; {\rm km}\; {\rm s}^{-1}  {\rm pc}^2)^{-1}$, 
 where the integrated intensity $I_0 = 300$ K km s$^{-1}$.
For $J=3-2$,  $X_{\rm CO(3-2)}  \simeq 2\times 10^{21}$ cm$^{-2} (I_{\rm CO(3-2)}/I_0)^{3.0}$  $({\rm K}\; {\rm km}\; {\rm s}^{-1})^{-1}$, or
$\alpha_{\rm CO(3-2)} \simeq 44 M_\odot \;  (I_{\rm CO(3-2)}/I_0\; )^{3.0}$  $({\rm K}\; {\rm km}\; {\rm s}^{-1}  {\rm pc}^2)^{-1}$, 
 where the integrated intensity $I_0 = 300$ K km s$^{-1}$. Here, we assume $v_{turb} =$ 10 km s$^{-1}$ and $ G_0= 1000$. 
For other fitting results, see Table 1\footnote{
{One may notice in the X-factor for CO(3-2) and CO(4-3) that  some fraction of  grid cells, where the intensity is less than
 $I_{\rm CO} \sim100  ({\rm K}\; {\rm km}\; {\rm s}^{-1})^{-1}$, 
do not follow the positive power law. This corresponds to the faint, optically thin outer edge of the disk (see Fig. 3).}}.
 
One should also note that the dispersion of $X_{\rm CO}$ 
is more than one order of magnitude for a given intensity, which reflects the fact that the physical and chemical conditions are far from uniform in the circumnuclear disk and are also due to the effect of the line transfer in the inhomogeneous medium (\S 3.1 and Fig. \ref{wada_fig: fig3}). The strong positive dependence of the X-factor on the intensity implies that the lines are saturated 
with $I_\nu = S_\nu$ in optically thick, high density clumps, which are more frequent along the line of sight in 
bright regions, as seen in Fig. \ref{wada_fig: fig3}a. These conditions in the ISM around the AGN 
are in contrast with the local GMCs or 
molecular gas in the disks of normal galaxies, where the X-factor is often assumed as roughly constant.

We found that the X-factor does not strongly depend on the value of FUV ($G_0$) but tends to be smaller for larger $v_{turb}$ (see {Fig. \ref{wada_fig: 9}e and  Fig. \ref{wada_fig: 9}f} and discussion in \S 4.2).

%

\begin{deluxetable*}{cccccc}
\tablenum{1}
\tablecaption{Coefficients and powers in fitting the X-factor distributions (Fig. \ref{wada_fig: 4}) with 
$X_{\rm CO} = $ $a \;10^{20}$ cm$^{-2} (I_{\rm CO}/I_0 \;)^{b}  ({\rm K}\; {\rm km}\; {\rm s}^{-1})^{-1}$,  \\ 
 $\alpha_{\rm CO} = c \; M_\odot \;  (I_{\rm CO}/I_0\; )^{d}  \; ({\rm K}\; {\rm km}\; {\rm s}^{-1}  {\rm pc}^2)^{-1}$. }
\tablewidth{12cm}
\tablehead{ \colhead{line} & \colhead{$a$} & \colhead{$b$} & \colhead{$c$} & \colhead{$d$} & \colhead{$I_{fit}$}  }
\startdata
1-0 & 2.0 & 4.8 & 4.4 & 4.8 & 600  \\
1-0 & 2.5 & 1.5 & 5.5 & 1.5 & 300  \\ \hline
2-1 & 15.8 & 7.4 & 34.4 & 7.4 & 600 \\ 
2-1 & 5.0 & 1.3 & 10.9 & 1.3 & 300 \\ \hline
3-2 & 20.0 & 3.0 & 44.0 & 3.0 & 300  \\
3-2 & 4.0 & 0.5 & 8.7 & 0.5 & 100 \\ \hline
4-3 & 200.0 & 4.3 & 436.0 & 4.3 & 300 \\
4-3 & 12.6 & 0.9 & 27.5 & 0.9 & 100 \\
\enddata
\tablecomments{$I_0 \equiv 300 ({\rm K}\; {\rm km}\; {\rm s}^{-1})^{-1}$. The conversion factor in the intensity range of $I  > I_{fit}$ is fitted.
}
\end{deluxetable*}

%
\subsection{The CO ladder}
%

\begin{figure}[h]
\centering
\includegraphics[width = 8cm]{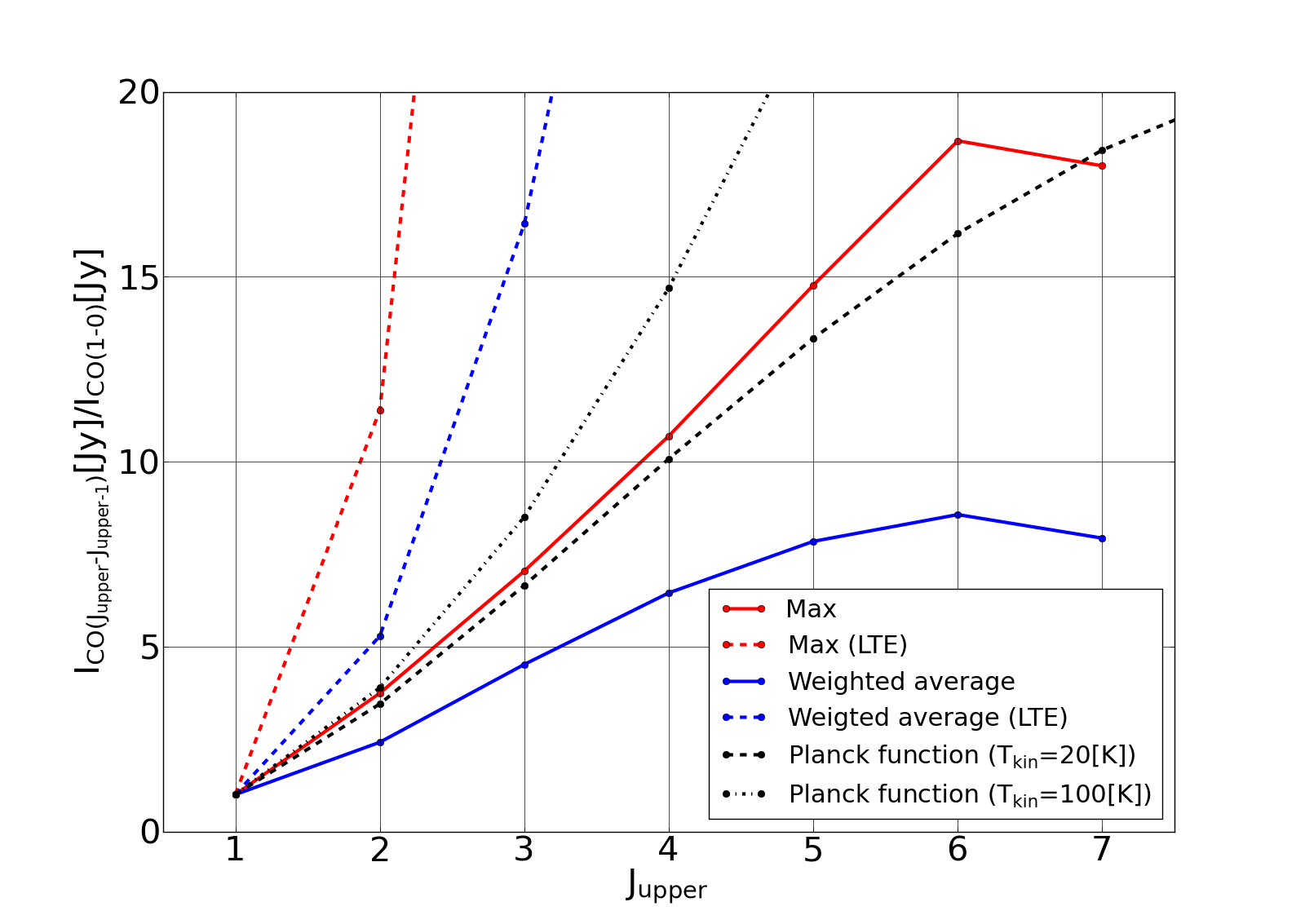}  
\caption{SLED of CO lines, normalized to the intensity of $J=1-0$ in Jy 
for the brightest grid point (red) and intensity weighted average of the whole area (blue). 
For comparison, the LTE cases are plotted as red and blue dashed lines, respectively. The black dotted and dashed lines 
are the line ratios for the blackbody of uniform gas with $T_{gas} =$ 100 and 20 K, respectively. }
\label{wada_fig: 5}
\end{figure}

In Fig. \ref{wada_fig: 5}, we show the total integrated intensity of the CO lines
as a function of the rotational
transition number $J$, i.e., the SLED for
the brightest spot and the intensity-weighted average value.
Both show that the intensity with respect to CO(1-0) has a peak around $J=6$.
The SLED of the brightest spot is close to that for the gas in 
 LTE with {$T_{gas} = 20$ K (black dashed line)}.
As shown by the weighted average (blue line) in the non-LTE case,  
most regions of the disk projected on the sky (see Fig. \ref{wada_fig: 2}) are not thermalized. 
This is because the internal structures of the disk are considerably clumpy and 
the line of sight is mostly occupied with optically thin gas (see Fig. \ref{wada_fig: fig3} and \S 3.1).

Note that if we assume LTE for all grid cells and calculate the line transfer, as for the non-LTE case,
 the resultant SLEDs (dashed blue and red lines) are very different.
 In a non-uniform medium with a large difference in optical depth, the LTE assumption in each grid cell
 does not necessarily cause the ``observed'' line ratios of the integrated intensities to be proportional to
 $\nu^2$. In the present case, the CO(1-0) becomes relatively weaker than CO(3-2) because of the line transfer
 effect; as a result, the line ratios appear 'super-thermal'.  

\citet{zhang2014} showed the CO SLED using Atacama Pathfinder EXperiment (APEX) for the central 18" ($\sim 360$ pc) of the Circinus galaxy and demonstrated that
the observed $^{12}$CO SLED has a maximum at $J=5-6$. Using the large velocity gradient (LVG) approximation, they suggested that 
the gas density of H$_2$ is $10^{2.7-3.8}$ cm$^{-3}$, the temperature is 80--400 K, and the velocity gradient $dv/dr$ is 1--25 km s$^{-1}$ (assuming 
a uniform abundance of CO of $x_{\rm CO} = 8 \times 10^{-5}$).
As the radius of the observed region is ten times larger than that of the present model, we cannot directly compare
our model with this APEX result. 
Moreover, the abundance distribution of CO should not be uniform as suggested by our result.
However, the suggested ranges of the physical conditions 
are consistent with those in the high-density gas in our numerical model. 

%

%
\subsection{Physical conditions of the ISM inferred from line ratios}
%
Line ratios are often used to infer the physical states (e.g., density, temperature, and velocity dispersion or gradient) of the ISM
in Galactic and extra-galactic objects.
For molecular lines, the numerical results with the LVG approximation or 
one-zone numerical code, such as RADEX \citep{vandertak2007}, can be compared with the observed line ratios,
taking the velocity gradient and fractional abundance as free parameters.
In our 3-D model, the physical conditions differ between grid cells on a scale of a few parsecs along the observed lines of sight; therefore, a specific observed line ratio does not necessarily represent a single physical state of the ISM.

\begin{figure}[h]
\centering
\includegraphics[width = 9cm]{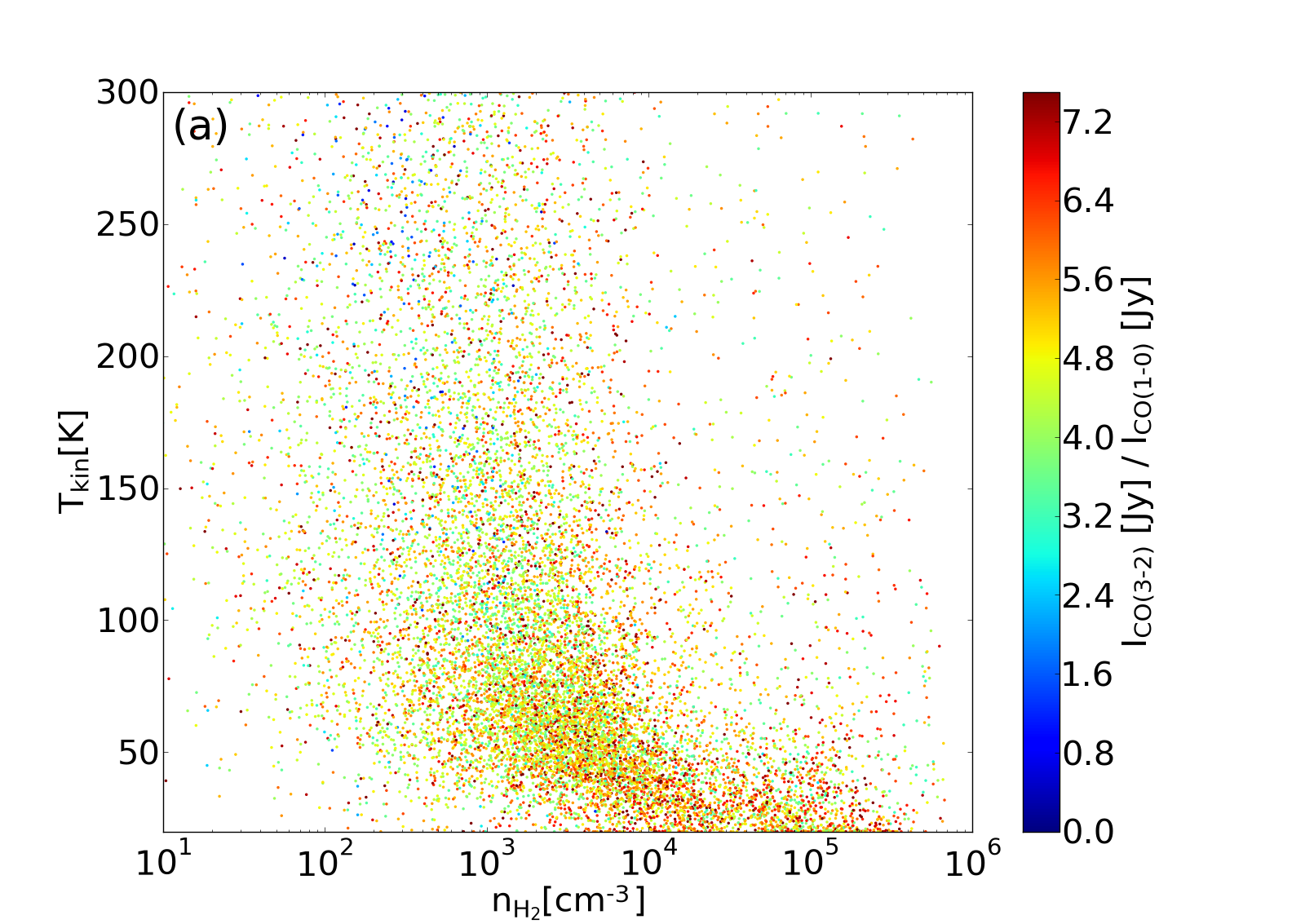} 
\includegraphics[width = 9cm]{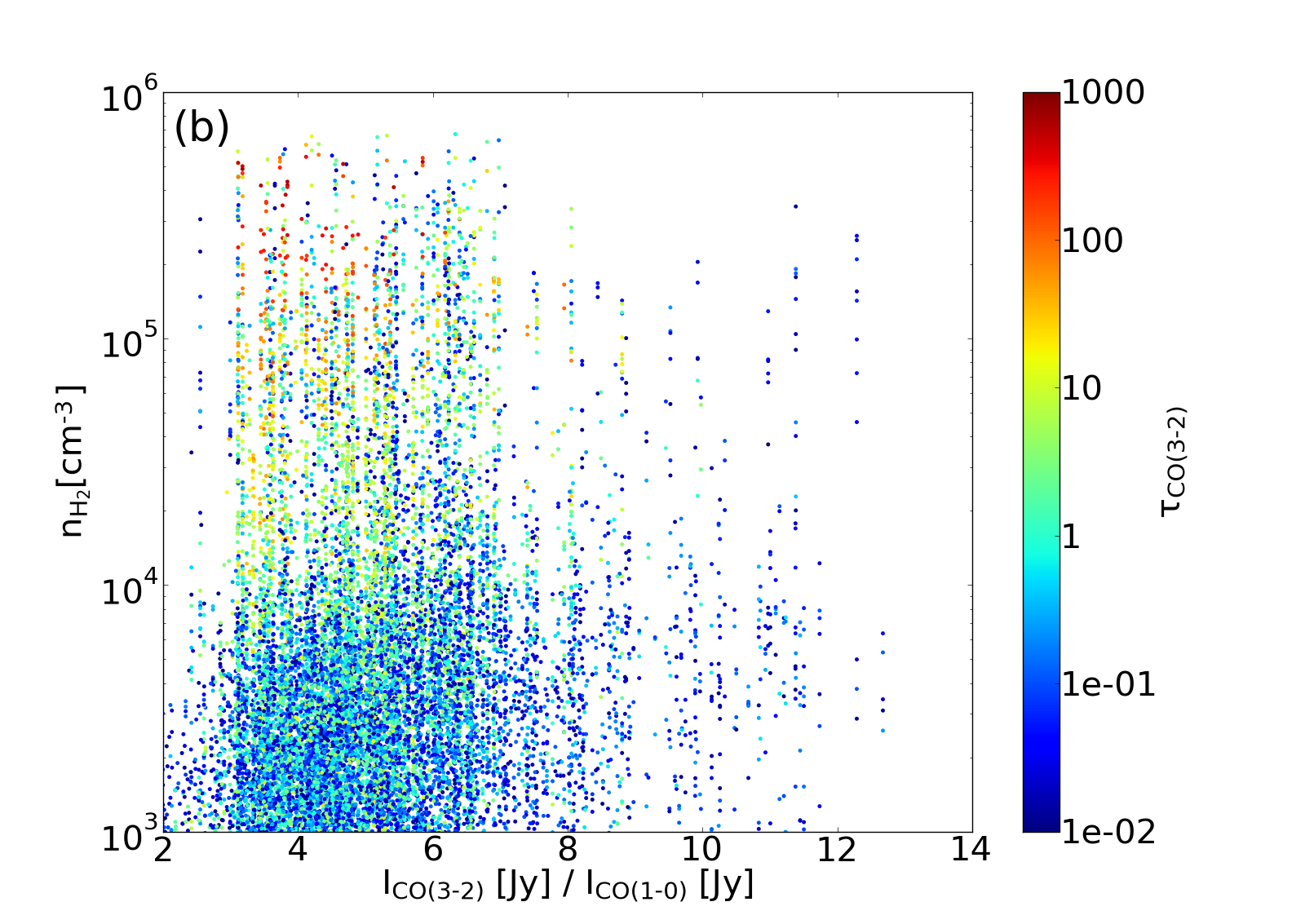} 
\caption{(a) Observed line ratios ({$I_{\rm CO(3-2)}/I_{\rm CO(1-0)}$}) in Jy  as a function of 
volume density and temperature in the grid cells along a line of sight.  Here, the viewing angle is assumed as edge-on, i.e., $\theta_v = 90^\circ$. (b) Similar to (a), but for the density of each grid cell. The color represents the optical depth of each 
grid.  }
\label{wada_fig: 6}
\end{figure}

In Fig. \ref{wada_fig: 6}a, the line ratio (in Jansky) of $R_{32} \equiv I_{\rm CO(3-2)/CO(1-0)}$ is plotted on a plane of the 
gas density and kinetic temperature derived from the input hydrodynamic data for 
the viewing angle of 90$^\circ$. Only grid cells with $\tau_\nu > 0.1$ are plotted.
It indicates that a line ratio could have been produced with more than three orders of magnitude in density.
There is a weak tendency that $R_{32} \sim 5$ for higher-density gas ($n \gtrsim 10^3$ cm$^{-3}$) and
$R_{32} \lesssim 3 $ for lower-density gas ($n \lesssim10^2$ cm$^{-3}$). However, 
there are significant exceptions. The gas temperature is almost independent of $R_{32}$, except for the highest-density gas ($n \gtrsim 10^4$ cm$^{-3}$), where $R_{32} \gtrsim 5 $ and $T_{kin} \lesssim 50$ K.

\begin{figure}[h]
\centering
\includegraphics[width = 9cm]{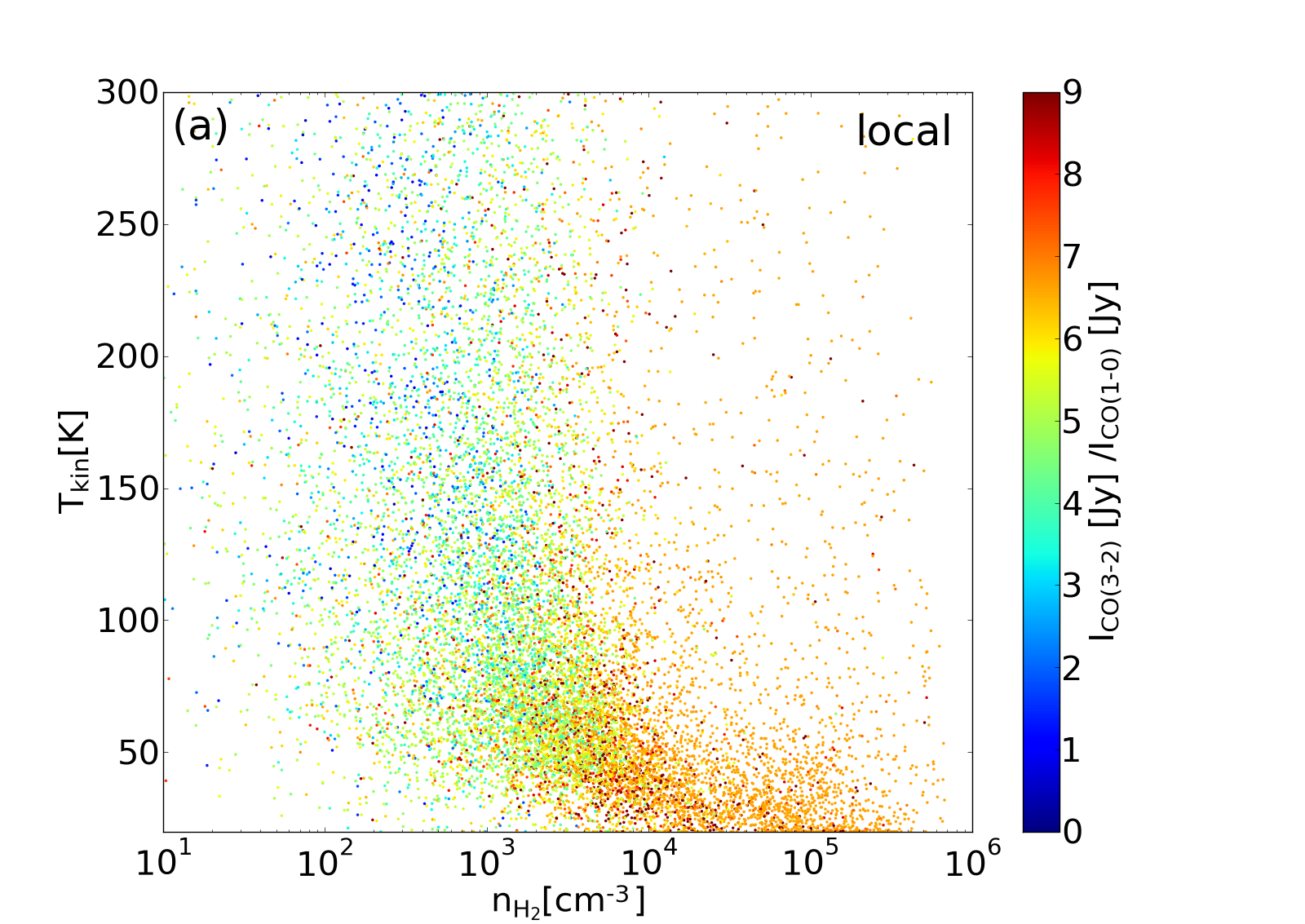}  
\includegraphics[width = 9cm]{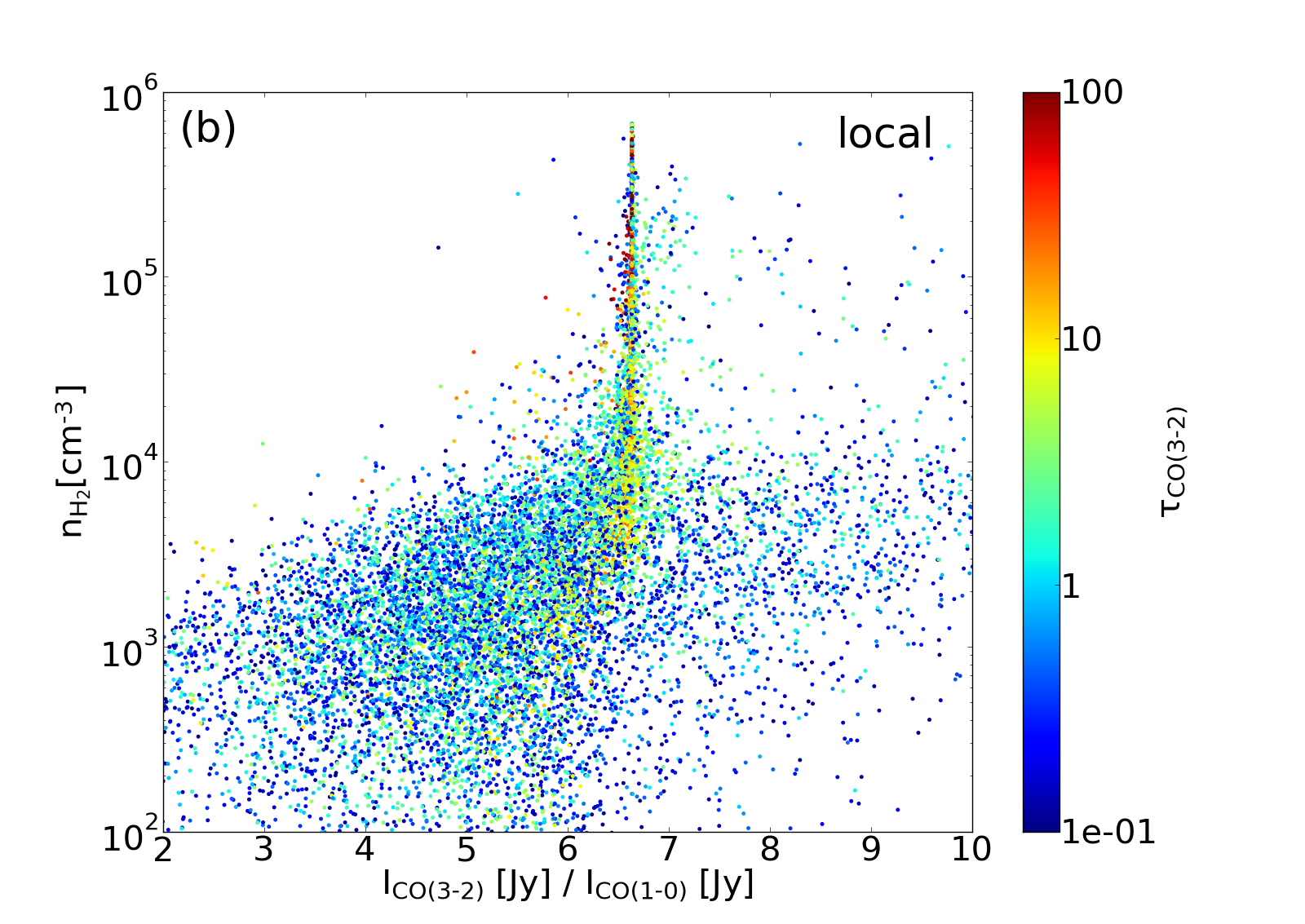}  
\caption{Same as Fig. \ref{wada_fig: 6}, but for local intensity calculated in each grid cell.}
\label{wada_fig: 6local}
\end{figure}

Figure \ref{wada_fig: 6}b illustrates how the optical depth of CO(3-2) depends on the H$_2$ density and
$R_{32}$. In the optically thick grid cells, the gas density is $n> 10^4$ cm$^{-3}$, where 
$R_{32} \sim 3-6$. Note that there are a small number of ``super-thermal'' grid cells where $R_{32} > 9$. 
These points do not necessarily indicate that the gas along its line-of-sight is excited non-thermally, 
 but it is caused by the difference in the line transfer effect in different lines.
Fig. \ref{wada_fig: 6local}a and b are, in contrast, plotted against {\it local} value of $R_{32}$ that is
the line transfer effect over grid cells is not taken account.
The local density now corresponds more clearly than in the non-LTE case.
Fig. \ref{wada_fig: 6local}b shows that $R_{32} \sim 6.5$ for optically thick grid cells ($\tau_{\rm CO3-2}  \gg 1$), 
 where $n_{{\rm H}_2 } > 10^4$ cm$^{-3}$ and $T_{gas} \sim 20$ K.
 The ratio is expected for the gas in LTE. 
 Most of the other grid cells are optically thin and not thermalized.

%

%

\newpage
%
\section{Discussion}
%

In our numerical modeling, there are two free parameters that could affect the results, i.e., 
the velocity dispersion inside one grid cell (i.e., micro-turbulence) and 
the strength of the FUV radiation. We discuss these effects below.

\subsection{Effect of micro-turbulence}

The molecular line profile in our treatment is assumed as a Gaussian profile with 
a dispersion ($v_{turb}$) originating from the unresolved internal turbulence in a grid cell. 
The relevant value of $v_{turb}$ in the molecular clouds in external galaxies, especially 
under the influence of the AGN, is still not clear.
One possible clue to estimate it is the size-dispersion relation of molecular clouds in 
the Galactic center. \citet{tsuboi2012} suggested, using CS(1-0), that the velocity 
dispersion of molecular gas in the
central molecular zone and in ``50 km s$^{-1}$ molecular clouds'' is about five times 
larger for a given size than that in typical GMCs in the Galactic disk.  This corresponds to $v_{turb} \sim 7 $ km s$^{-1}$ for
our grid size, i.e., 0.25 pc. As the nuclear region of the Circinus galaxy should be disturbed by
intense feedback from the AGN, as well as from starbursts, in contrast to our Galactic center,
the velocity dispersion on that scale could be larger than 10 km s$^{-1}$. 

\begin{figure}[h]
\centering
\includegraphics[width = 8cm]{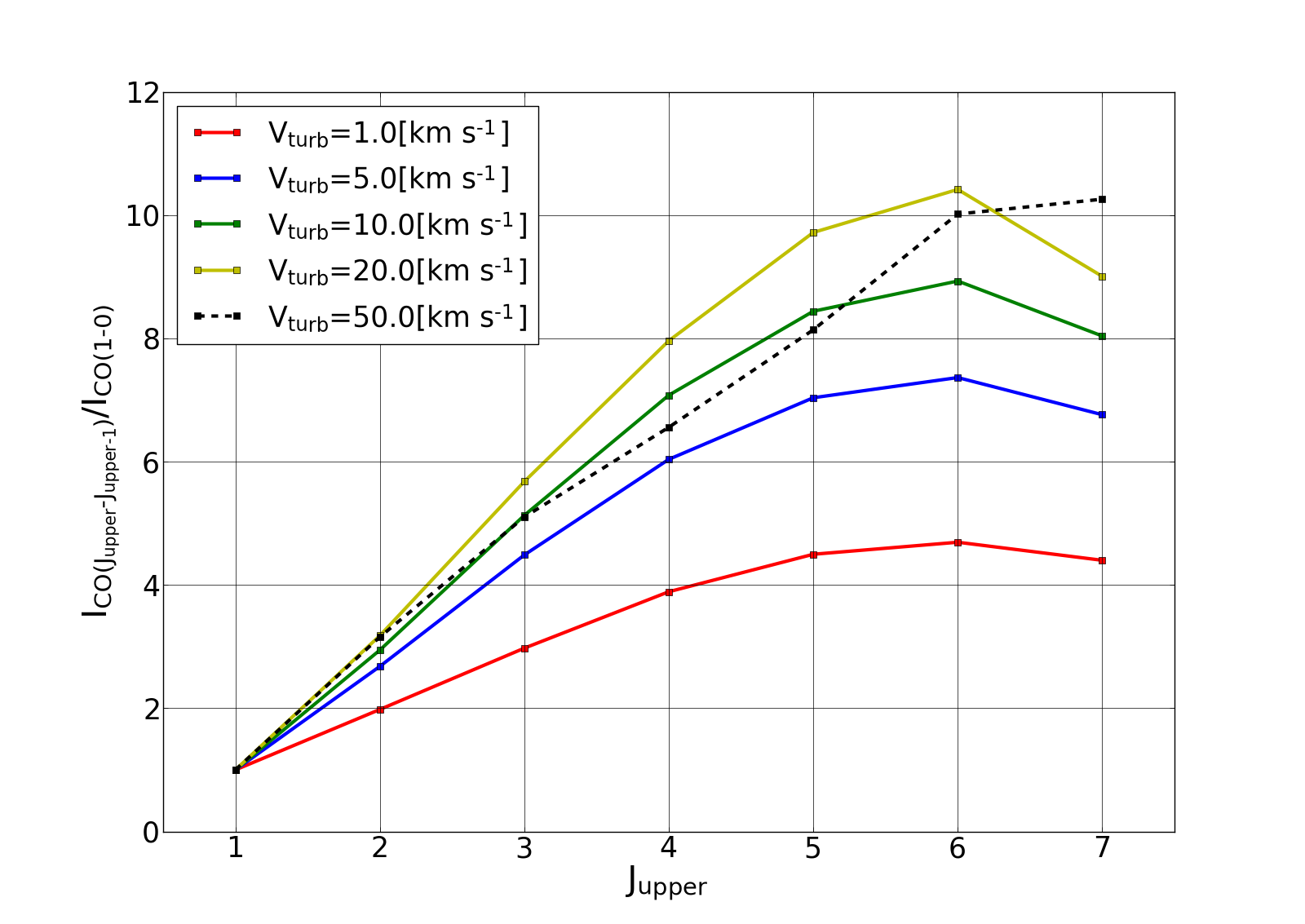}  
\caption{{SLED for five models with different micro-turbulence $v_{turb}$, i.e. unresolved turbulent motion inside one grid cell. }　}
\label{wada_fig: 7}
\end{figure}

\begin{figure}[h]
\centering 
\includegraphics[width = 8cm]{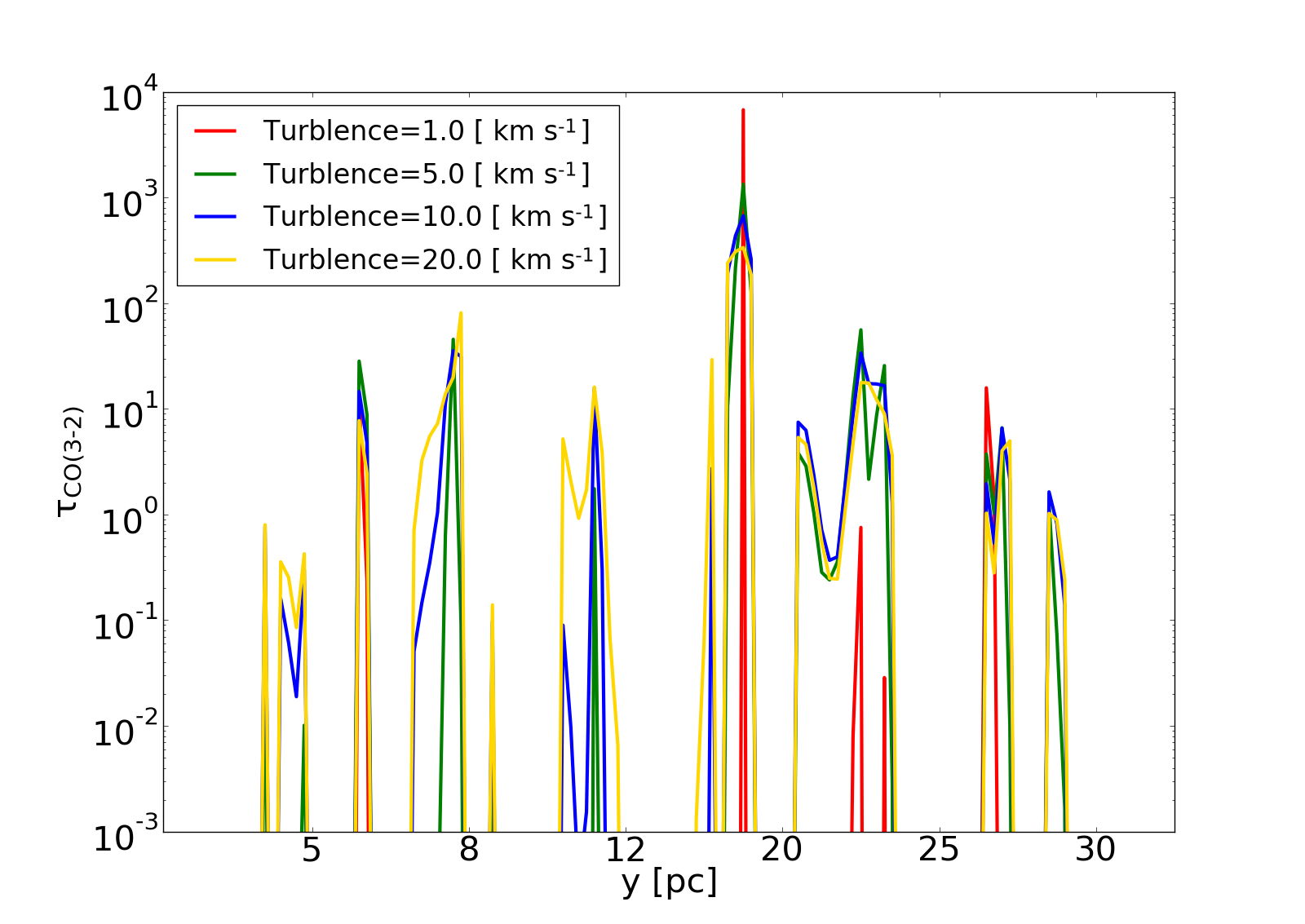}   
\caption{Optical depth (CO(3-2)) distribution along a line-of-sight in four different models with varying internal turbulent velocities $v_{turb}$. }
\label{wada_fig: 8}
\end{figure}

Figure \ref{wada_fig: 7} shows the SLED, similar to Fig. \ref{wada_fig: 5}, but for {the micro-turbulence $v_{turb} = $1, 5, 10, 20 and 50 km s$^{-1}$.}
Although the behavior of the intensity in terms of $J$ is similarly independent of $v_{turb}$, 
it shows that the intensity for a given $J$ increases with $v_{turb}$.
This is because the optical depth is proportional to the line profile function $\phi_\nu$.
Figure \ref{wada_fig: 8} presents the optical depth distribution for a given line of sight.
It shows that for larger $v_{turb}$, the intensity at the line center decreases,
but the total integrated intensity increases because there are more regions with $\tau_\nu > 0$
\footnote{{If  $v_{turb}$  is as large as $\sim 50$ km s$^{-1}$, the total intensity for $J \le 6$ turns to decrease because of the 
insufficient intensity at the line center as shown in the black dotted line in Fig. 9.}}.
The present results suggest that 
if we obtain the SLED in observations with a sufficiently fine spatial resolution (e.g., sub-parsec) for 
the central tens parsecs in nearby AGNs, 
we could determine a relevant value of the micro-turbulence in the molecular clouds.
This will be discussed in a subsequent paper (Izumi et al. in preparation). 

\begin{figure}[h]
\centering
\includegraphics[width = 12cm]{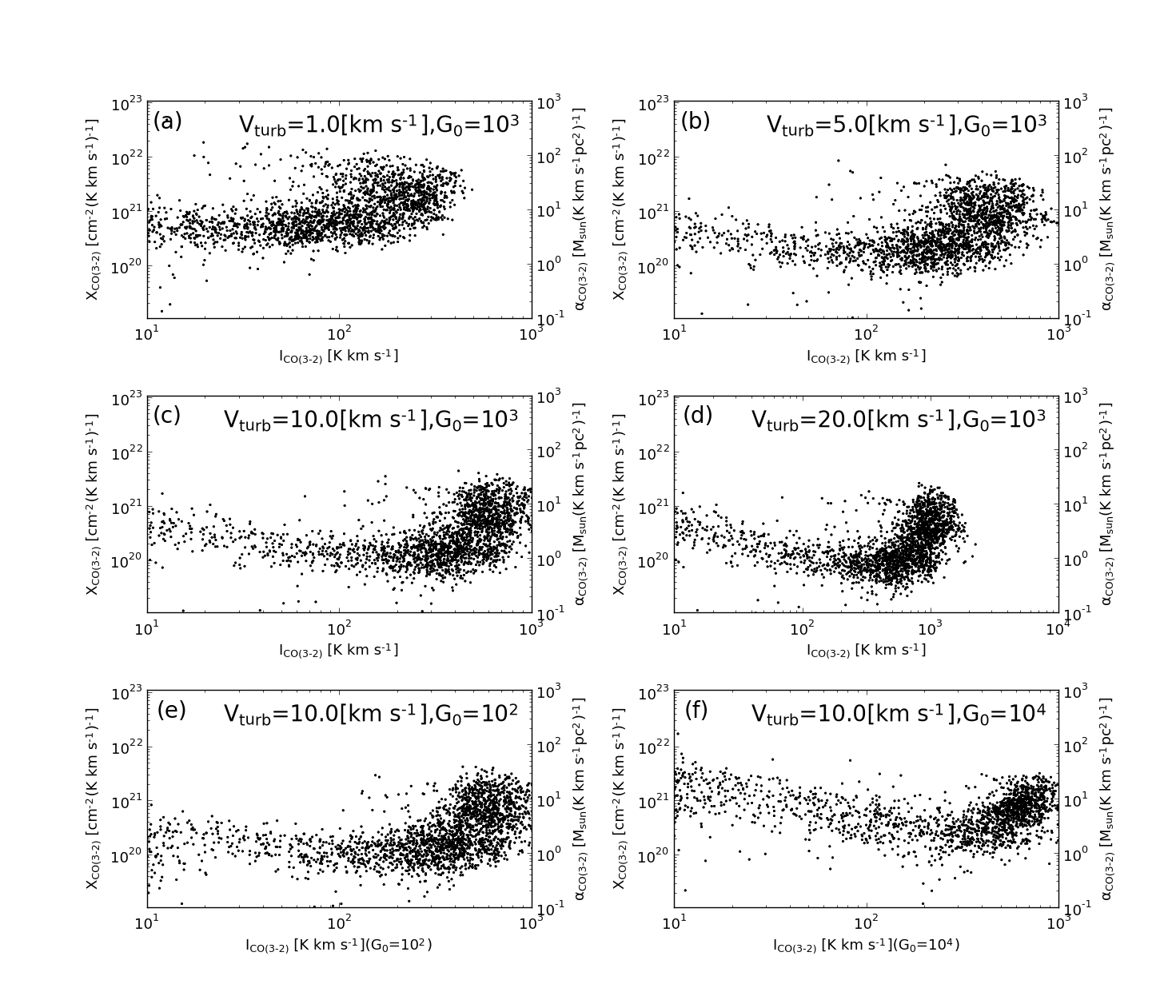}     
\caption{{Same as Fig. 5, but for six models with different turbulent velocities of 1, 5, 10 and 20 km s$^{-1}$  (a-d) and with different FUV of 
$G_0 = 100$ and $10^4$ (e and f).}}
\label{wada_fig: 9}
\end{figure}

{Figure \ref{wada_fig: 9}a-d} show the conversion factor ($X_{\rm CO(3-2)}$) for different values of micro-turbulence (1, 5, 10, and 
20 km s$^{-1}$), showing that the X-factor tends to be smaller for
larger micro-turbulence values for a given line intensity.
However, the dispersion does not significantly change among the results.

%

%

%

\subsection{Effect of FUV}

FUV field strength is one of the important quantities to determine the chemistry and thermal structures in the photo-dissociation region around galactic central regions \citep{meijerink05, wada09}. In starburst regions, the local value of the FUV should vary over several orders of magnitude \citep[][]{rosenberg2014}, e.g., $G_0 =100-10^5 $, 
depending on the local structures of the ISM and the distance to the radiation sources. 
Assuming the observed star formation rate (SFR) in the Circinus galaxy \citep{hicks09}, i.e., $\sim 100 M_\odot$ yr$^{-1}$ kpc$^{-2}$,
and the relation between $G_0$ and SFR density in the IR-luminous merger NGC 1614 \citep{saito2017}, the
FUV in Circinus can be estimated as $G_0 = 10^{2}-10^{3}$ on average (note that the scatter could be extremely large, ranging from $G_0=10$ to $10^6$).

If we directly solve the radiative transfer equations for the FUV in the inhomogeneous media with many
radiation sources, our present model could be more realistic.
However, this is beyond our current numerical treatment, so we 
assumed here a uniform $G_0$ and varied it as a free parameter.
We confirm that the X-factor does not significantly 
depend on the strength of the uniform FUV between $G_0= 10^2$ and $10^4$ {(Fig. 11e and 11f)}, 
because both the intensity of CO and the molecular hydrogen density are similarly affected by the
change in FUV \citep{wada09}.

\begin{figure}[h]
\centering

\includegraphics[width = 14cm]{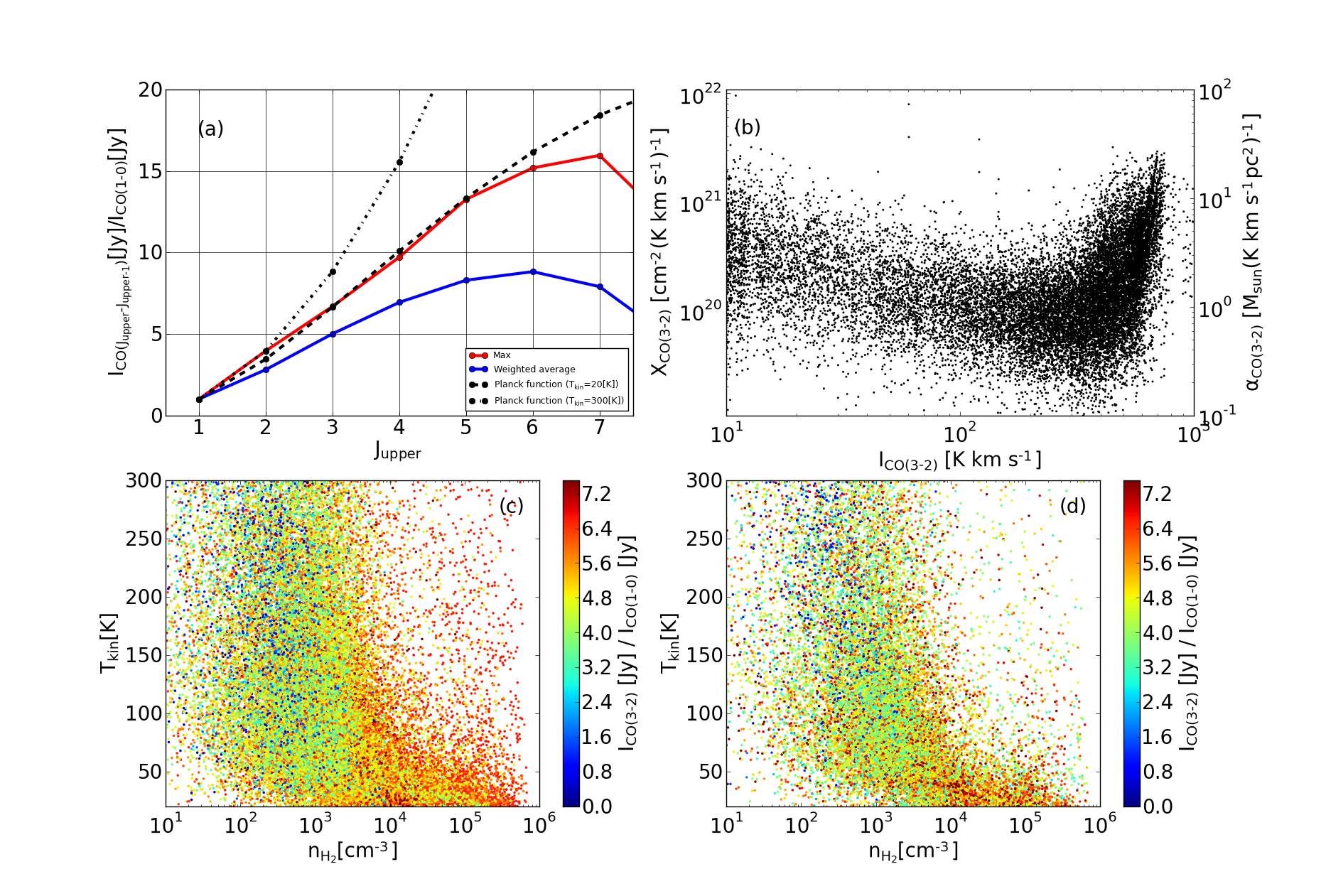}     
\caption{{Same as Figs. 5, 6, and 7, but for $\theta_v = 0$ (face-on).  (a) SLED, (b) X-factor for CO(3-2),  (c) $R_{32}$ vs. temperature and density.  (d) $R_{32}$ for the edge-on view. }}
\label{wada_fig: 12}
\end{figure}

{
\subsection{Effect of the Viewing Angle and Implications to Other AGNs}
So far we have focused  on the case of edge-on view, which is plausible for the circunumnuclear disk in Circinus \citep{wada2016}.
However, it would be worth to see how the results change for different viewing angles,
 in order to see whether the present results is applicable to other AGNs. 
Figure \ref{wada_fig: 12} shows the SLED, X-factor and $R_{32}$, in which the viewing angle $\theta_v  = 0$ (i.e. face-on) is assumed. 
The SLED (Fig. 12a)  can be compared with Fig. 6. The brightest spot (red solid line) in both cases close the LTE case with $T_{kin} = 20$ K, 
however, the intensity in the face-on case tends to be smaller, especially for high $J$. The weighted average value (blue solid line) does not
depend on the viewing angle.  The X-factor for CO (3-2) (Fig. 12b) can be compared with Fig. 5. There is no essential difference in terms of the
viewing angle. The line ratio of CO (3-2) ($R_{32}$) as a function of $T_{kin}$ and $n_{{\rm H}_2}$ for the face-on view is plotted in Fig. 12c. 
Grid cells with $\tau > 0.01$ are plotted. 
For comparison, the same plot for the edge-on case is shown in Fig. 12d. 
Although the line ratio tends to be larger for a given density in the face-on case, 
as discussed in \S 3.4, 
it also exhibits a large dispersion in gas density and temperature for a given line ratio.
This is because the molecular gas forms a thick disk (Fig. 1a), in which there is internal structure  on sub-pc scales along the $z$-axis. 
If this inhomogeneous nature of the ISM in the central tens parsecs exists in other AGNs, the results presented here should be 
also considered to interpreted CO line observations.}

{
Until now, the only `direct' detection of  molecular lines from circumnuclear disk (CND) with a spatial resolution of a few pc is NGC 1068,
which is the prototypical Seyfert 2 galaxy at $D = 14$ Mpc \citep{garciaburillo2016}. 
Using CO(6-5) line with ALMA,  they claimed that the CND is a 7-10 pc diameter disk and it should be a
counter part of the `dusty torus' suggested by the near IR and mid IR interferometric observations \citep[e.g.][]{jaffe04}.
Although the spatial resolution ($\sim$ 4 pc) is not fine enough to resolve sub-pc inhomogeneous structures, if they actually exist as found in our model 
for Circinus, the large velocity dispersion ($\sim 30 $ km s$^{-1}$)  and the lopsided distribution (i.e. $m=1$ mode) imply the presence of complicated
internal sub-structures. In fact, the observed SED  is well fitted by a `clumpy' torus model \citep{garciaburillo2016}. 
However, our model disk for Circinus is globally axisymmetric, and  no lopsided distribution of the molecular gas is seen in contrast to NGC 1068.
The condition for triggering the large scale asymmetry is an interesting subject to be explored  in terms of the hydro and magneto-hydrodynamic
instabilities, such as Papaloizou-Pringle  instability  \citep{pploizou84} and magneto-rotational instability.}
{The CO(6-5) observation of NGC 1068  shows no significant CO counterpart for
the polar emission of the dust tori, which are often seen in nearby AGNs \citep{lopez2016}.
This is also the case in our 'radiation-driven' fountain, where there is no dense molecular gas in the bipolar outflows. 
}

{The origin of sub-kpc to kpc scale molecular outflows/jets, such as in 
NGC 1068 \citep{garciaburillo2014}, Circinus \citep{zshaechner2016}, 
and NGC 1377 \citep{aalto2016},  
and its physical link to the cicumnuclear molecular gas
are still not clear. The presence of 10-pc scale 
molecular outflows in NGC 1068 is under debate, even if there is some 
distorted kinematics of  molecular gas in the torus is observed\citep{gallimore2016}.
This is an interesting subject to be explored by high-resolution
ALMA data of the central region of the Circinus galaxy, and will be discussed
in a subsequent paper (Izumi et al. in prep.). }

{Line ratios, such as CO(2-1)/(1-0) and CO(3-2)/(1-0) in the central 100 pc of  NGC1068 is
2-3  \citep{viti2014}, which is larger by a factor of two than those in our case (Fig. 3). 
On the other hand, \citet{zhang2014} showed that the ratio CO(3-2)/(1-0) in the central several 100 pc of
the Circinus galaxy is  about 0.8, which is consistent with our model. 
In the Circinus galaxy, there is no currently available high spatial resolution data for CO(1-0). 
Therefore, with these limited observational information at this moment and due to
the difference of the spatial resolution between models and observations,
it is hard to conclude on the consistency or inconsistency of the predicted line ratios with the observationally derived values.}

%

%
\section{Conclusions}
%
Recent observational and theoretical studies of obscuring material from sub-parsec to 100-pc scales around AGNs 
suggested non-static, multi-phase structures over wide density and temperature ranges.
In this study, we calculated the line emissions from cold, molecular gas, which 
was expected in our previous "radiation-driven 
fountain model" \citep{wada2016}.  
Using a snapshot of the 3-D radiation-hydrodynamic simulation that
reliably explains the SED of the nearest 
type 2 Seyfert galaxy, the Circinus galaxy, with XDR chemistry as 
an input, we conducted post-processed, 3-D, non-LTE line transfer simulations for CO lines ($J < 15$) in the central $r <$ 16 pc. 
We found that the CO emissions ($J=1-6$) mostly originated from thick, inhomogeneous structures.
There are almost no ``molecular outflows,'' as the outflowing gas density is too low and mostly ionized.
 The SLED has
 a peak around $J =  6$ and its distribution suggests that the lines are optically thin for most regions.
However, for a given line-of-sight, the optical depth distribution is highly non-uniform between $\tau_\nu \ll 1 $ and $\tau_\nu \gg 1$.

Because we know the molecular hydrogen density at each grid cell in the input model, 
we can obtain the CO-to-H$_2$ conversion factor ($X_{\rm CO}$ or $\alpha_{\rm CO}$).
We found that the conversion factor
depends strongly on the integrated intensity for a given line-of-sight, especially for lower $J$ lines:
\begin{itemize}
\item  $X_{\rm CO(1-0)}  \simeq 2.0\times 10^{20}$ cm$^{-2} (I_{\rm CO(1-0)}/I_0)^{4.8}$  $({\rm K}\; {\rm km}\; {\rm s}^{-1})^{-1}$, or
$\alpha_{\rm CO} \simeq 4.4 M_\odot \;  (I_{\rm CO(1-0)}/I_0\; )^{4.8}$  $({\rm K}\; {\rm km}\; {\rm s}^{-1}  {\rm pc}^2)^{-1}$, 
 where the integrated intensity $I_0 = 300$ K km s$^{-1}$.
 \item $X_{\rm CO(3-2)}  \simeq 2.0\times 10^{21}$ cm$^{-2} (I_{\rm CO(3-2)}/I_0)^{3.0}$  $({\rm K}\; {\rm km}\; {\rm s}^{-1})^{-1}$, or
$\alpha_{\rm CO} \simeq 44 M_\odot \;  (I_{\rm CO(4-3)}/I_0\; )^{3.0}$  $({\rm K}\; {\rm km}\; {\rm s}^{-1}  {\rm pc}^2)^{-1}$, 
 where the integrated intensity $I_0 = 300$ K km s$^{-1}$
\end{itemize}
One should note that there is large (more than one order of magnitude) scatter around this average value, reflecting the non-uniform internal structure (density, temperature, abundance, and velocity) of the ``torus.'' 
 We also found that the conversion factor for a given intensity 
 depends on the assumed value of the ``micro-turbulence,'' which is the
 velocity dispersion in one grid cell (0.25 pc in the present case). 
The values above could be the case for the central region of the Circinus galaxy, 
 and they could also be applied to low-luminosity AGNs associated with
 nuclear starbursts, but not necessarily to all types of AGNs, such as luminous quasars.
 
 The total CO intensities depend on the assumption of {`micro-turbulence, i.e. unresolved velocity dispersion} inside a grid cell (0.25 pc).
 It is brighter for larger velocity dispersions (Fig. \ref{wada_fig: 7}), provided that $v_{turb} \lesssim 20$ km s$^{-1}$.
Using this fact, we can estimate the internal turbulent motion of the circumnuclear gas in Circinus
 and compare it with ALMA Cycle-4 observations of the central molecular disk (Izumi et al. in prep.).
  
   We also found that  the physical conditions differ between grid cells on a scale of a few parsecs along the observed lines of sight; therefore, a specific observed line ratio,
   such as $I_{{\rm CO}(3-2)}/I_{{\rm CO}(1-0)}$,   does not necessarily represent a single physical state of the ISM.
The dense ISM ($n > 10^4$ cm$^{-3}$) on a 0.25-pc scale is mostly in a phase of molecules and LTE, but the 
resultant line ratio for an observer is not necessarily in LTE, because of the line transfer effect.
{These results basically do not depend on the choice of the viewing angle, therefore
what we found here could be useful to understand physics of the ISM around supermassive BHs,
not only in the Circinus galaxy, but also in other nearby AGNs.}
The present results also suggest that we need to carefully analyze the molecular line observations
 of the circumnuclear gas when we obtain them with a high spatial resolution in nearby AGNs by ALMA.

\acknowledgments
The authors are grateful to the anonymous referee for his/her constructive comments and suggestions. 
We also thank
L. Hunt, M. Schartmann, and A. Burkert for their fruitful comments and suggestions. KW thanks the Max Planck Institute for Extraterrestrial Physics and Excellence Cluster Universe for their
support in 2017. Numerical computations were performed on a Cray XC30 at the Center for Computational Astrophysics at the National Astronomical Observatory of Japan. This work was supported by JSPS KAKENHI Grant Number 16H03959 and 17K14247.

\end{document}